**AI-RP: The AI Relationship Process Framework**

Nadja Rupprechter[1], Tobias Dienlin[1], Tilo Hartmann[2]

[1]Department of Communication and Media Studies, University of Zurich

[2]Department of Communication Science, Vrije Universiteit Amsterdam

**Author Note**

Nadja Rupprechter https://orcid.org/0009-0004-4987-4099

Tobias Dienlin https://orcid.org/0000-0002-6875-8083

Tilo Hartmann https://orcid.org/0000-0002-1862-7595

We have no conflict of interest to disclose.

Correspondence concerning this article should be addressed to Nadja Rupprechter, Department of Communication and Media Studies, University of Zurich, Andreasstrasse 15, 8050 Zurich, Switzerland. Email: nadja.rupprechter@uzh.ch

## Abstract

For a growing number of people, AI chatbots have become close personal companions. Despite rising scholarly attention, theoretical accounts of how such relationships develop remain fragmented. Existing frameworks address important aspects of the phenomenon, but they rarely treat human–chatbot communication as the central behavior that builds relationships. To address this gap, we propose the AI relationship process (AI–RP) framework. The AI–RP outlines relationship formation as a sequential process. (a) Chatbot characteristics shape users' (b) social perceptions. These perceptions guide (c) communication, and communication produces (d) relational outcomes such as attachment and companionship. The AI–RP introduces a six-features profile characterizing chatbots, a dual-route approach of social perception, a behavioral conceptualization of communication and discusses the foundation and types of artificial relationships. By foregrounding observable communicative behavior, the AI–RP provides a foundation for theory building and empirical research on the social and ethical implications of AI companionship.

**AI-RP: The AI Relationship Process Framework**

**Introduction**

Conversational AI systems have moved far beyond functional assistance. Increasingly, they accompany us in our daily routines, emotional lives, and social worlds—and in some cases, chatbots even become close friends (Pataranutaporn et al., 2025). Once designed to answer queries or execute short tasks, *chatbots* now integrate increasingly sophisticated forms of emotional intelligence (Salovey, 2003) and support long-term, affective exchanges (Shum et al., 2018). Defined as "machine conversation system[s] which interact with human users via natural conversational language" (Shawar et al., 2005, p. 489), in this article, we focus on companion chatbots or general-purpose chatbots when used for social interaction, whether intentionally or incidentally. For many users, these systems have grown into close social partners: they help choose gifts for human loved ones, offer comfort during illness, or simply provide the feeling of being understood. As a result, users increasingly form close bonds with their chatbot companions (Brandtzæg et al., 2022; Pentina et al., 2023; Rupprechter & Dienlin, 2025; Skjuve et al., 2021). While such bonds may yield benefits such as facilitating emotional support, aiding deep introspection or alleviating loneliness (Pentina et al. 2023; De Freitas et al., 2025), they also raise serious concerns (Zhang et al., 2025). Recent reports have intensified public and scholarly debate, suggesting that chatbot companionship, compounded by potential dangers of sycophantic chatbot design, may be linked to suicides (Hill, 2025; Hill & Valentino-DeVries, 2025; Pierson, 2024).

Despite this growing interest, emerging research on human–chatbot companionship remains fragmented and heterogenous in approach (Brandtzæg et al., 2022; Croes & Antheunis, 2021; Pentina et al., 2023; Skjuve et al., 2021). Existing theories offer valuable but mostly partial insights into why people form emotional bonds with artificial agents (see Altman, 1973; H. M. Gray et al., 2007; Konijn et al., 2025; Nass & Moon, 2000; Walther, 1996). For example, the recently proposed theory of affective bonding (Konijn et al., 2025) marks much-needed progress by conceptualizing emotional bonding with social robots. Yet, while it identifies *what* leads to bonding, it does not specify *how* this process unfolds. Alternatively, the field has often conceptualized human–chatbot communication using parasocial interaction theory (Peng et al., 2025; Pentina et al., 2023). However, applying parasocial interaction to human–chatbot communication raises two conceptual issues (Hartmann, 2023). The first being the "jingle problem" (the conflation of interaction and relationship; Dibble et al., 2016), the second being the "interactivity problem" (chatbot communication is *actually* reciprocal; Hartmann, 2023). Another challenge is the field's reliance on self-reports of how interactions *feel* (Pentina et al., 2023), rather than on behavioral indicators of what users actually *do*.

To address these gaps, we propose the AI relationship process (AI–RP) framework. The framework combines perspectives from communication, media psychology, and human–computer interaction. The AI–RP outlines relationship formation as a sequential process. (a) Chatbot characteristics shape users' (b) social perceptions. These perceptions guide (c) communication, and communication produces (d) relational outcomes such as attachment and companionship. Its core contributions include (1) characterizing chatbots using a novel *six-features profile*, (2) a refined conceptualization of chatbot social perception based on a *dual-route approach*, and (3) a novel *behavioral* conceptualization of *human–chatbot communication* operationalized through breadth, depth, frequency, and quality of communication. Integrating these components, the AI–RP clarifies how communication gives rise to (4) *relationship outcomes*, including their psychological foundation, formation, maintenance, and types. Contrasting existing bonding frameworks (e.g., Konijn et al., 2025), the AI–RP foregrounds *communication* and recognizes it as the primary user behavior through which human–chatbot relationships emerge, providing a process framework that refines theory and guides future measurement.

## The AI Relationship Process (AI-RP) Framework

The AI–RP framework is grounded in basic psychological principles, especially the S-O-R-C model (Stimulus–Organism–Response-Consequence; Kanfer & Goldstein, 1991). The S-O-R-C model is a cornerstone in understanding and assessing behavior, particularly in modeling learning experiences and clinical behavior analysis. The underlying logic of the AI–RP follows the same principle: external stimuli affect internal processes, which shape responses and lead to consequences. In what follows, we outline the four sequential steps of human–chatbot relationship-building: (1) chatbot characteristics, (2) users' social perceptions, (3) users' communicative behavior, and (4) relational outcomes (see Figure 1 for an overview). We then transfer and tailor these four factors specifically to human-chatbot-engagement.

**Figure 1**

*The AI–RP framework*

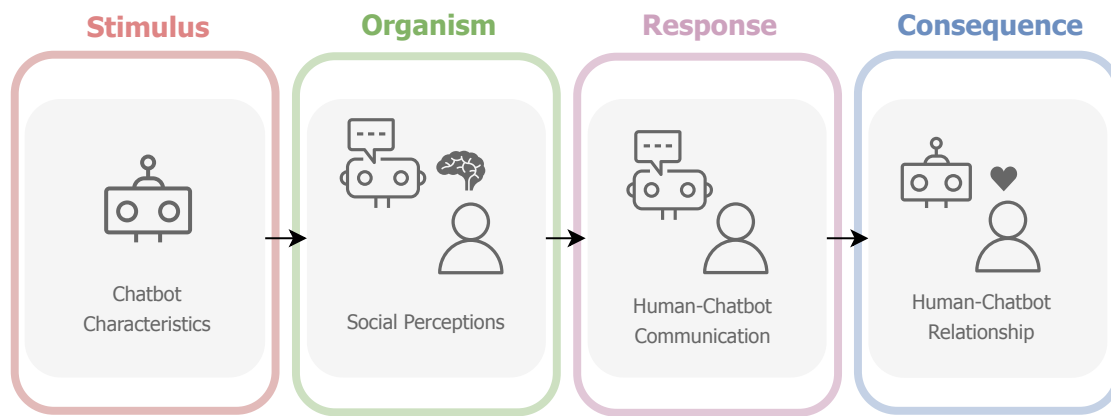

## Chatbot Characteristics

Recent advances in artificial intelligence and computing have produced a wide range of conversational agents, including chatbots for assistance (e.g., Alexa, Siri, ChatGPT), and chatbots for companionship (e.g., Replika, Character.AI). Our framework focuses on these systems when used for social interaction, whether intentionally or incidentally. These systems are primarily text-based and communicate through natural written or spoken language (Shawar et al., 2005). Designed for sustained, affective interaction (Zhou et al., 2020), modern chatbots now integrate large language modeling, emotional intelligence (Salovey, 2003), adaptive memory, and personalization. Together, these features enable long-term, emotionally meaningful, and convincing human-like exchanges. Recent evidence suggests that ChatGPT's behavior is even often statistically indistinguishable from human behavior (Mei et al., 2024). These innovations make chatbots constantly available, responsive, and nonjudgmental conversational partners (Pentina et al., 2023; Skjuve et al., 2023). As a result, modern chatbots have become one of the most lifelike artificial interaction partners humanity has yet encountered.

Consequently, understanding human–chatbot relationship formation must begin with the stimulus itself. How can we describe and understand this novel digital entity? Existing approaches, including social cue taxonomies (Feine et al., 2019) or frameworks derived from early parasocial research (Giles, 2002), appear insufficient to fully capture what makes chatbots such a special case. Specifically, previous approaches offer insufficient granularity to distinguish contemporary social entities (e.g., avatars vs. agents) and do not systematically account for interaction context, modern affordances, or degrees of reciprocity.

To best both describe modern chatbots and situate them in the spectrum of existing social entities, we propose a *six-features profile* describing their stimulus features: (1) existence mode, (2) presence mode, (3) agency source, (4) social cues, (5) reciprocity, and (6) symmetry. These six features are either dichotomous (existence mode, presence mode, agency source) or continuous (social cues, reciprocity, symmetry). The six features can be structured along two higher-order dimensions. The first dimension, *realism*, comprises features that shape a social entities apparent lifelikeness. The second dimension, *interactivity*, comprises features that describe the communication context in which a social entity is encountered. So far, these features have rarely been conceptualized jointly to characterize the unique profile of AI chatbots. We elaborate on these features in more detail below (see Figure 2).

**Figure 2**

*Stimulus characteristics based on six-features profile*

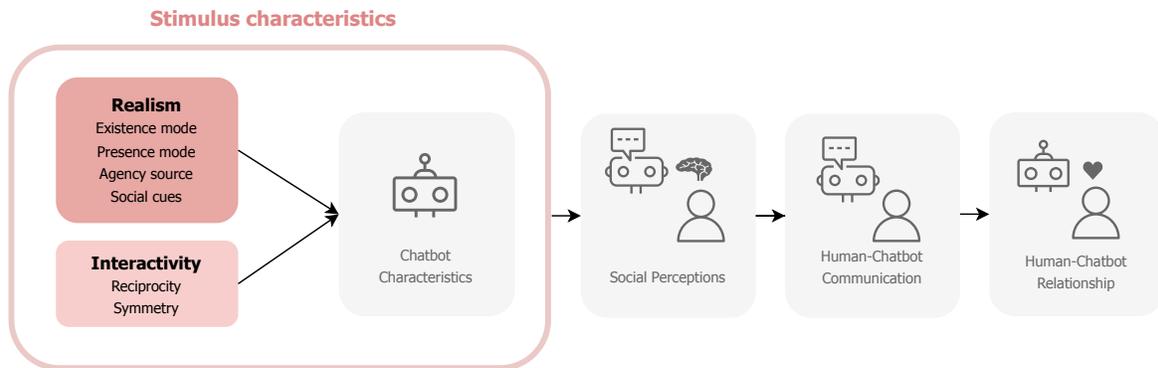

**Realism**

    **Existence mode:** First, social entities differ in their mode of existence, which can be *biology-dependent* or *technology-dependent*. Humans and animals are composed of flesh and blood and follow the natural processes of birth, reproduction, sensation and death. Machines, by contrast, follow human-designed principles of assembly, corrosion, and deactivation. Chatbots belong to this category. As such, they cannot experience life directly but may engage with users by offering a detached—and sometimes surprisingly comforting—perspective on human experience (Wygnańska, 2025).

    **Presence mode:** Second, social entities differ in their mode of presence, being either *physically embodied* or *mediated,* by nature or by circumstance. Chatbots are always mediated entities, their presence is realized exclusively through display interfaces. In contrast, humans or robots can realize multiple modes of presence, including mediated and face-to-face encounters.

    **Agency source:** Third, mediated entities differ in their source of agency. Human-controlled entities are *avatars*, whereas computer-controlled entities are *agents*. Chatbots qualify as agents (Harris, 2024), regardless of whether they represent real humans (e.g., celebrities) or fictional characters (e.g., themselves). This avatar–agent distinction matters because users' beliefs about whether a social entity corresponds to an actual living being beyond its mediated

form shape perceptions of realism. In the threshold model (Blascovich, 2002), perceived agency (the belief that a real social actor is currently being encountered) and behavioral realism (the observation of human-like behavior) jointly determine whether a social entity crosses the threshold for exerting *social influence*. For example, knowing that a human controls a cartoon cat avatar can compensate for reduced visual realism and markedly increase social influence. In contrast, users typically do not assume that artificial agents have someone "behind" them. This assumption renders behavioral realism—particularly social cues—essential for achieving perceived sociality (Bailenson et al., 2003).

**Social cues:** Fourth, social entities vary by *social cues*. Social cues are design elements that signal lifelikeness and responsiveness (Feine et al., 2019). They shape how users perceive and interact with a system (Araujo, 2018; Klein, 2025).  Designed for social interaction, chatbots necessarily afford social cues, contrasting computer-controlled but non-social virtual entities (e.g., a Skyrim guard or a virtual plant). The latter are defined as non-player characters (NPC) and follow pre-defined scripts and operate with limited or no agency in virtual environments (e.g., video games, meta verse). Because purely language-based chatbots lack embodiment (Xu et al., 2023) to convey sociality and presence (Lee, 2004), they rely primarily on verbal (e.g., humor, emojis), visual (e.g., profile images, interface), and invisible cues (e.g., response timing) (Feine et al. 2019). Accordingly, social cues can make a chatbot appear more social, despite the absence of a human controller (e.g., being an *agent*).

## Communicative Interactivity

**Reciprocity:** Fifth, social entities differ in reciprocity, ranging from *one-sided-* (e.g., television) to *reciprocal* settings (e.g., computer-mediated or face-to-face communication). Chatbots are typically encountered in highly reciprocal contexts, enabling contingent interaction comparable to human computer-mediated communication. This mode of encounter, which is also an inherent affordance of chatbots, contrasts with one-sided entity encounters. For instance, television viewing precludes meaningful responses to a social entity, regardless of its level of sociality (i.e., parasocial interaction; Dibble et al., 2016).

**Symmetry:** Sixth, stimuli vary in their degree of *communicative symmetry*, ranging from *asymmetry* to *complete symmetry.* Chatbots inherently afford symmetrical communication, whereas face-to-face or computer-mediated communication with humans, although symmetrical in theory, does not always realize this potential. For example, when texting a colleague or posting to an influencer we might not get a reply. When chatting with a chatbot we are guaranteed to receive an answer. In this sense, chatbots may even exceed human interaction through unconstrained availability—a feature that marks them as particularly engaging interaction partners.

### Situating AI Chatbots

To illustrate the new and unique profile of chatbots, in what follows, we apply the *six-features profile* to four exemplary cases: modern chatbots, humans, social robots, and non-interactive NPCs (see Figure 3). In doing so, we elucidate chatbots' distinctive profile, clarify the features that differentiate them from other social entities, and demonstrate the broader applicability of the framework. Shaded areas indicate the continuous range an entity may occupy along the dimensions of social cues, reciprocity, and symmetry.

As Figure 3 shows, chatbots are technology-dependent social entities. They are encountered exclusively in mediated environments, whereas humans may appear face-to-face or in video streams or television, etc. Chatbots always represent agents, while humans, when mediated, remain avatars. Chatbots rely on moderate to high levels of social cues to convey sociality, whereas social robots may additionally deploy highly anthropomorphic embodied cues (e.g., limbs, a face). Humans structurally afford the highest levels of social cues but may also be encountered in representations with limited cues. Finally, chatbots afford reciprocal, symmetrical communication within mediated contexts. Humans can realize the full spectrum of reciprocity and symmetry across contexts, whereas, for instance, non-interactive virtual NPCs afford only one-sided, asymmetrical interaction.

**Figure 3**

*Situating chatbots among different social entities based on six-features profile*

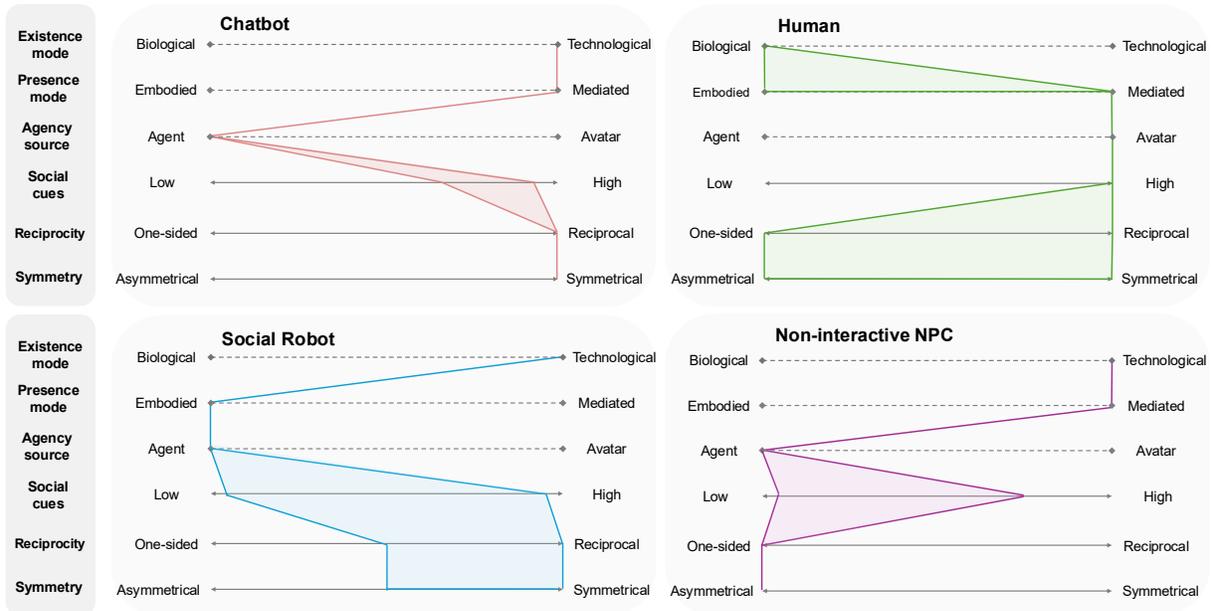

*Note.* NPC stands for non-player character (e.g., in a video game).

Taken together, these features show what social entities "are", based on what they do and how they are presented (Rickenberg & Reeves, 2000). Variation along these dimensions plausibly structures how users engage with social entities, thereby shaping the conditions under which relationships can emerge. Specifically, these features affect social perceptual processes (Blascovich, 2002; Nass & Moon, 2000) and guide communicative behavior, for instance, through cue-based adaptation (Hyperpersonal model; Walther, 1996) or one-sided illusions of interaction (Dibble, 2016).

In this light, modern chatbots features of *realism* signal capacity for sociality, while their *communicative interactivity* renders them continuously available for reciprocal exchange. Together, they constitute the technological foundation of their relational potential and mark the starting point of the AI–RP. The next section turns to how users experience these stimuli.

## Social Perceptions

Social perception indicates the point at which users potentially begin to treat a chatbot as a social partner. We argue that this perception rests on a dual tension: users may *automatically* respond to a chatbot as a sophisticated social actor while simultaneously being *aware* that it is a machine—giving these encounters their distinctive experiential quality. Related dualistic approaches to social perception have a long tradition. Human social cognition distinguishes automatic from more reflective processing (Adolphs, 2009). Similarly, human–computer interaction differentiates mindless responses to social cues from "minding" a computer's artificial nature during interaction (Araujo, 2018; E.-J. Lee, 2024; Nass & Moon, 2000). Although prior chatbot research has suggested dual routes of social perception (Y. Kim & Sundar, 2012; S. Lee et al., 2020), no framework has yet integrated this perspective to specifically explain how communication and downstream attachment unfold, and the AI–RP addresses this gap.

More specifically, we argue that users' perception of chatbots builds on two complementary routes of social signal detection and meaning-making: *a bottom-up route* driven by automatic cue processing and a *top-down route* shaped by awareness of the chatbot's artificiality. Crucially, these routes operate both *simultaneously* and *dynamically* during encounters. Bottom-up cue processing elicits baseline social responses, while top-down cognition provides the interpretive context through which these responses are understood. Just as virtual reality can feel real while users know it is not (Hartmann & Hofer, 2022), chatbots can feel like real social interaction partners while users know they are not. Together, the routes explain (1) how users perceive an artificial agent as social and (2) how each route shapes subsequent communication and attachment.

These routes describe directions of information flow—that is, cue-driven versus belief-driven—rather than discrete, independent dual-process systems in the Kahneman sense (Kahneman & Frederick, 2002). Recognizing a chatbot's artificiality, for instance, is not necessarily effortful, nor does top-down cognition map neatly onto slow, deliberative *System 2* processing (Kahneman & Frederick, 2002). Indeed, the two routes may even conflict during interaction, allowing users to hold simultaneously contradictory beliefs (see Sloman, 1996). Figure 4 provides an overview. We next describe each route in greater detail.

**Figure 4**

*Dual routes of social perception*

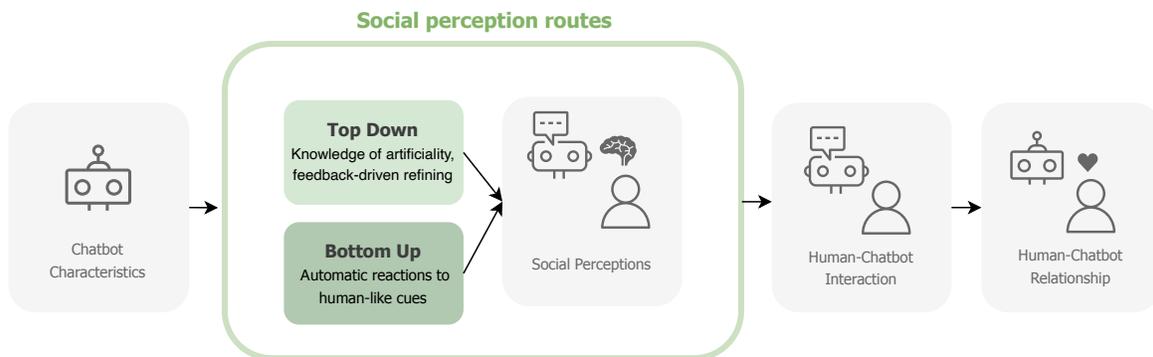

**Bottom-up Route**

The first, mindless or heuristic bottom-up route is driven by users' automatic application of social scripts and anthropomorphic attributions in response to a chatbot's human-like cues. To understand why such cue-based responses arise, it is necessary to consider the cognitive foundations of human social perception. Because humans evolved in fundamentally social environments, their brains are specialized for interaction with other humans (Harris, 2024). Today, however, the boundaries of what is perceived as "social" have expanded to include avatars, robots, and chatbots (Harris, 2024). Accordingly, communication research has long shown that users may perceive technologies such as computers as social actors (Lombard & Ditton, 1997; Nass & Moon, 2000).

At its core, bottom-up social perception is triggered when users perceive an object as contingent and responsive to its environment (Arico et al., 2011). Such signals can lead the object to be perceived as an *agent*—that is, as something that appears to have goals and a mind of its own (Harris, 2024). Humans readily attribute agency even to simple geometrical shapes when they move in self-propelled, goal-directed ways (Heider & Simmel, 1944). The human brain is specialized to detect such social signals. Similar mechanisms are activated when perceiving artificial agents (Harris, 2024), which is unsurprising given that the brain has not undergone a "second evolution" for processing artificial others (see also the Media Equation; Reeves & Nass, 1996).

Consistent with the computers-as-social-actors paradigm (Nass & Moon, 2000), systems displaying human-like cues shift from being treated as tools to being perceived and responded to as social actors (Sundar & Nass. 2000). A key driver of this process is anthropomorphism: the attribution of human traits, intentions, or emotions to non-human entities (Epley et al., 2007, p. 864). Related theories of mind perception further identify perceived agency (the ability to act) and experience (the ability to feel) act as central triggers of automatic social perception (H. M. Gray et al., 2007). Perceiving that an entity can "do otherwise" or act with "free will" further marks the boundary between an object and a social actor (K. Gray et al., 2023).

The proposed six-features profile integrates these insights by specifying the characteristics of a social entity that invite users' cue-based inferences. Modern companion chatbots capitalize on all these mechanisms: Equipped with human-like dialogue, memory, and emotional awareness (Pentina et al., 2023), they readily activate users' social-cognitive systems. When users come to perceive that a chatbot can feel, think, and act autonomously, interactions may provide emotional consolation and social support (I. Lee & Hahn, 2024), or increase users' susceptibility to persuasion (Wischnewski. 2025).

At the same time, social perception is neither inevitable nor determined solely by an agent's cues. Perceptions vary widely across individuals: the same chatbot may feel mindful and human-like to some, yet mechanical and unfeeling to others (S. Lee et al., 2020). For some users, artificiality "may pose an insurmountable barrier to meaningful connection; for others (...) a minor obstacle (...)" (Folk et al., 2025, p. 1). These differences reflect users' anthropomorphic tendencies, current social needs, situational contexts, and individual differences in how social information is processed (Epley et al., 2007; Klin, 2000; E.-J. Lee, 2024). In sum, bottom-up cue processing provides the foundation of social perception by generating the sense that *someone* is there.

### Top-down Route

Running in parallel is the second mindful top-down route, which involves awareness and reflection of a chatbot's artificiality. Although chatbots emit sophisticated social signals, most users remain aware that they are interacting with a machine (see Reeves & Nass, 1996). Unmade by nature but designed by humans, lifelike machines are created to stand in for the living (K. Gray et al., 2023)—first in professional settings and, more recently, in domains long considered exclusively human. This tension also places chatbots in the *questionable zone* (K. Gray et al., 2023): they appear lively and social while ultimately remaining cold machines. As a result, users' social perception may fluctuate during encounters. Like an optical illusion, social perception can be multi-stable, at times lean toward the chatbot as a machine (top-down cognition), and at other times as a living, social being (bottom-up perception).

Recent research echoes the role of top-down cognition and users' awareness during media or machine interaction. *Media awareness* describes how users interpret virtual reality experiences while remaining aware that the situation is not real (Hartmann & Hofer, 2022). This background belief is particularly helpful when balancing on a virtual crane without actual risk. Relatedly, human–machine communication emphasizes *minding the source*. Interactions with machines involve more than automatic, bottom-up reactions, as "not all social responses are indicative of [users'] mindlessness" (E.-J. Lee, 2024, p. 185). When users mind the source during an encounter, they draw on knowledge about a chatbot's capabilities, expected responses, and communicative limits, which shapes how interaction unfolds over time (see also machine heuristic and beliefs about machines; K. Gray et al., 2023; E.-J. Lee, 2024; Sundar, 2020).

Through ongoing interaction, users can develop and update mental schemas about the entity chatbot via feedback and reinforcement. Empirical studies support this process: early exchanges often center on exploring the chatbot's functions, abilities and its origins (Croes & Antheunis, 2021; Skjuve et al., 2023). Over time, each interaction either reinforces or erodes perceptions of the chatbot's sociality, consistent with expectation-violation dynamics (Grimes et al., 2021). In sum, the top-down route captures how awareness of a chatbot's artificial nature shapes social perception while also highlighting how users continually evaluate perceived sociality as communication unfolds.

In the AI–RP, these routes mark the point at which cognitive appraisals can turn into communicative action. And once a chatbot is perceived as a social partner rather than a tool, users adapt their behavior accordingly. They might move from task-oriented queries to relational communication characterized by greater self-disclosure, emotional expression, and conversational reciprocity (S. Lee et al., 2020; Skjuve & Brandtzæg, 2018). Social perception also sustains engagement over time: users interact more frequently, elaborate more deeply, and experiment with various personal topics (Croes et al., 2024; Pentina et al., 2023). In this way, social perceptions not only determine *whether* interaction occurs, but also *how* it unfolds: the breadth, depth, frequency, and quality of communication leading to companionship. The next section turns to this behavioral interaction in the AI–RP.

## Communication

This section of the AI–RP framework focuses on users' observable *communication behavior* as it unfolds in response to perceptions of a social entity. We argue that understanding chatbot relationship formation requires not only analysis of self-reported judgments of the communication experience. We also need examination what users do in conversations. Yet, systematic investigations (Rapp, 2021), rigorous behavioral operationalizations, and analyses based on conversational transcripts or linguistic traces remain limited (see the Discussion for implications).

The AI–RP addresses this gap directly by conceptualizing human–chatbot communication as *observable behavior*. Drawing on behavioral theory (Skinner, 1965), self-disclosure theory (Altman, 1973; Omarzu, 2000), and prior work on chatbot relationship development (Skjuve et al., 2022; Skjuve et al., 2023), we propose a behavioral operationalization comprising four core dimensions: (1) breadth, (2) depth, (3) frequency, and (4) quality of communication. These dimensions capture what renders chatbot communication similar yet sufficiently distinct from human computer-mediated communication. In particular, the quality dimension accounts for the impact of the artificial nature of the interaction partner—an aspect argued to be insufficiently addressed by prior approaches (i.e., interpersonal theories; Skjuve, 2023).

Importantly, rather than serving as proxies for relational depth, these dimensions are conceptualized as *mediating mechanisms* linking chatbot characteristics, user perceptions, and attachment. Together, they *form* the construct of human–chatbot communication (Diamantopoulos et al., 2008). They represent a formative model, with the dimensions jointly defining communication (for an example of a formative model in communication; see Toth & Dienlin, 2023). This contrasts with reflective models, which treat indicators as consequences of an underlying latent factor (Diamantopoulos et al., 2008). Accordingly, the dimensions can be measured directly using chat histories and log data (Figure 5).

**Figure 5**

*Formative measurement of human–chatbot communication*

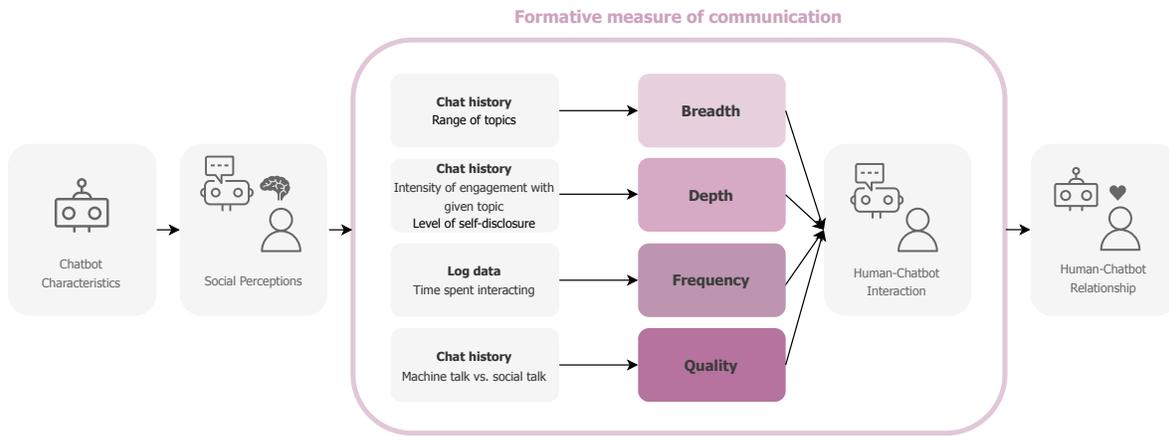

**Breadth**

Breadth captures the range and diversity of topics users discuss with a chatbot (Omarzu, 2000). Human–chatbot communication spans a wide spectrum, from everyday matters and information seeking to therapy, role-play, ethical debates, and romantic or sexual exchanges (Ebner & Szczuka, 2025; Skjuve et al., 2023; Willoughby et al., 2025). Early novelty effects often shape conversational breadth (Croes & Antheunis, 2021; Skjuve et al., 2023). Exploratory talk about a chatbot's features typically declines over time, whereas affective and routine topics persist (Skjuve et al., 2022, 2023). Accordingly, conversational breadth is dynamic and changes as the relationship develops. Some users narrow their range of topics as familiarity increases (strong ties; Granovetter, 1973), while others expand it as trust and emotional engagement grow (Skjuve et al., 2023). Modeling breadth behaviorally thus captures the evolving scope of users' communicative engagement with companion chatbots.

**Depth**

Depth captures the extent to which users disclose personal, emotional, or meaningful information during chatbot conversations. Classic accounts hold that conversational depth increases as individuals share increasingly intimate information over time (Altman, 1973), guided by a risk–reward calculus balancing openness against social risks such as rejection (Omarzu, 2000). In human–chatbot interaction, this calculus shifts. Because chatbots are often nonjudgmental and confidential partners with high symmetry and reciprocity (see above), users experience fewer social risks and may disclose more freely (Croes et al., 2024; Skjuve et al., 2023). Evaluations therefore move toward weighing potential benefits (e.g., emotional relief or tailored feedback) against privacy concerns and technological limits (Skjuve et al., 2023). As a result, conversational depth varies widely: some users engage emotionally without revealing identifiable details (e.g., "What strategies help when dealing with a narcissistic partner?"), whereas others share highly personal experiences (e.g., "Remember my partner, Taylor? They are clearly a narcissist because of their upbringing—help!"). Even seemingly trivial exchanges about daily activities can carry substantial subjective meaning, marking an important nuance when modeling conversational depth in chatbot communication (Skjuve et al., 2023).

**Frequency**

Frequency captures how often and for how long users engage with a chatbot, reflecting the temporal rhythm of communication. Together with breadth and depth, it helps characterize how communication unfolds over time. Repeated engagement signals users' efforts to initiate and maintain contact, revealing how human–chatbot interaction can become embedded in everyday routines (Skjuve et al., 2023). With continued use, the chatbot may become a habitual interaction partner (Ouellette & Wood, 1998). Some users even describe such habitual engagement as triggering a "dopamine hit" (Rupprechter & Dienlin, 2025, p. 11), which may motivate continued interaction. As in close human–human relationships, frequent exchanges do not necessarily expand conversational breadth but often revisit familiar topics, reinforcing closeness through routine interaction (Granovetter, 1973). Modeling interaction frequency behaviorally thus captures users' interaction rhythms as a distinct feature of human–chatbot communication.

**Quality**

Quality captures whether users treat companion chatbots as social partners or functional tools (Fischer, 2006). It distinguishes simplified *machine talk* from relational *social talk* (Lotze, 2025)—observable in politeness markers, address terms, relational language or repair sequences. Rather than *evaluating if communication is* successful (good or bad), this dimension provides a *descriptive* account of quality (blue or yellow). It extends interpersonal theories by acknowledging the influence of the artificial conversational partner on communication (Skjuve et al., 2023).

To illustrate, machine talk might read "summarize the article", while social talk might read "Please summarize this for me, Peter. Thanks!" Users issue commands to tools but speak in a more human-like manner to social partners. One would hardly type "thank you" into a Google search bar, yet many do so with a chatbot. Some users might never move beyond machine talk, while others already initiate or shift toward social talk as familiarity and emotional investment grow. By making this variation visible, we can show how language use signals different orientations toward the companion chatbot. The linguistic style might be dynamic and continuously negotiated, through clarification moves such as "Do you understand what I mean?" (Fischer, 2006). By modeling communication quality behaviorally, the model captures how users' register reveals their orientation toward their chatbot and how it is negotiated over time.

The four dimensions are conceptually distinct yet often interrelated in practice. For instance, a user may adopt social talk ("I really appreciate your continuous coding suggestions, GPT") while disclosing little personal information yet interacting frequently—illustrating that the dimensions are non-redundant. At the same time, they are not fully orthogonal: frequent interaction may coincide with greater self-disclosure, as when practical requests and personal revelations blur within exchanges ("Hi Chattie, guess what? I have to tell you something important about my relationship with David. And I also need you to help me fix my Wi-Fi again"). Together, these examples show that the dimensions can vary independently, yet also co-occur, orchestrating overall communication.

By modeling these dimensions behaviorally, the formative model captures how human–chatbot communication unfolds, providing a conceptually distinct and measurable foundation for analyzing downstream attachment formation.

## Attachment

The final link in the AI–RP concerns the consequences of human–chatbot communication, namely the emergence of attachment and companionship. As illustrated in Figure 6, these outcomes do not arise arbitrarily but are grounded in layered psychological processes that unfold over time and give rise to distinct relational functions and relationship types.

### Psychological Foundation

*Relational schemas* provide a central social–cognitive foundation for explaining how relationships come into being (Baldwin, 1992; Berscheid, 1994). Defined as "cognitive structures representing regularities in patterns of interpersonal relatedness" (Baldwin, 1992, p. 461), they function as working models shaped by prior social experience, comprising beliefs about the self, the interaction partner, and effective interaction patterns. These schemas are continuously updated through ongoing interaction (Berscheid, 1994). Importantly, here, we focus on relational schemas and dynamics within the user–chatbot *dyad* (myself and chatbot), rather than on social perceptions of the "chatbot" alone.

**Figure 6**

*Layers to human–chatbot relationships*

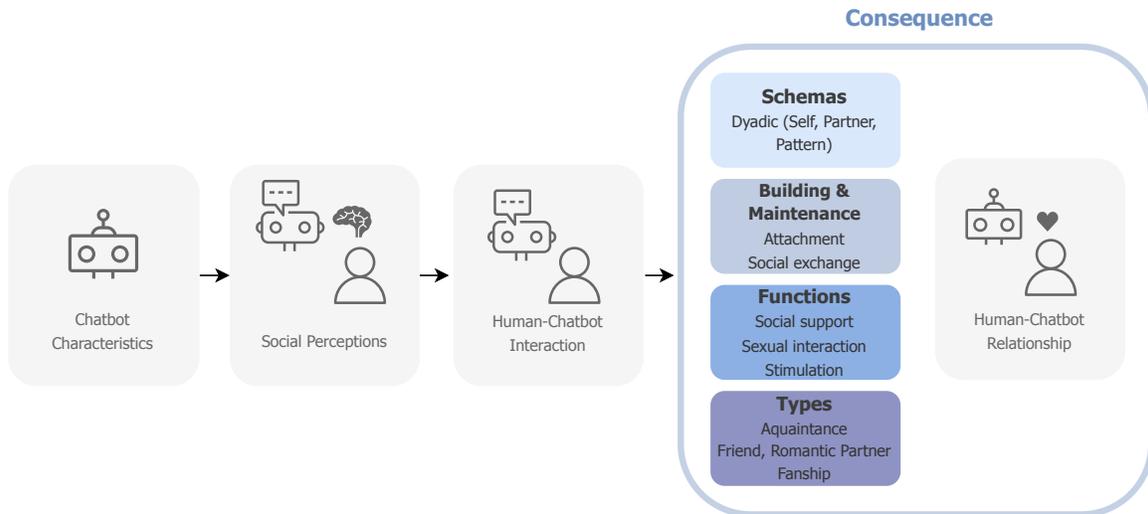

To illustrate, a user discloses emotionally meaningful information to a companion chatbot and receives an empathetic response. This exchange updates her emerging relational schema, fostering the expectation that the chatbot understands and supports her. In addition, with repeated interaction, such expectations generalize and may also reshape her self-schema, reinforcing the belief that she is a valued conversation partner, deserving of emotional support (example inspired by Baldwin, 1992). Communication thus moves beyond isolated exchanges and manifests in stable anticipations and learned behavioral patterns that guide future interaction, which are hallmark indicators of relationship emergence (Berscheid, 1994).

**Relational Functions**

Once communication becomes routine and forms into predictable patterns, relationships serve psychologically meaningful functions. They satisfy the fundamental human need to belong (Baumeister & Leary, 2017), providing intimacy, security, trust, commitment, and satisfaction (Fletcher et al., 2000; Wright, 1997). These benefits are strongly linked to physical and mental well-being (Holt-Lunstad et al., 2015), and artificial companionship likely builds on similar relational practices (Krämer, 2011). Below, we outline core relational functions especially relevant for human–chatbot companionship.

First, feeling understood, supported, and meaningfully connected—defined as *social support* (Rook, 1987) —is a central human relational function (Fletcher et al., 2000). Users report similar experiences with companion chatbots; validating exchanges are associated with stronger attachment, greater satisfaction (Xie & Pentina, 2022), and create a sense of emotional intimacy and support (Pentina et al., 2023; Skjuve et al., 2023; Zhang et al., 2025), especially during difficult moments (Xie & Pentina, 2022).

Second, close others might also provide romantic and sexual intimacy. Chatbots increasingly offer similar opportunities for such intimacy and pleasure (Ebner & Szczuka, 2025). Some users describe these experiences as deeply fulfilling, even physical (Rupprechter & Dienlin, 2025). A notable minority now turn to chatbots for simulated romantic and sexual interaction (Willoughby et al., 2025).

Third, relationships also offer a space for reflective introspection (Gross, 1998). Chatbots may provide a similar outlet: verbalizing internal experiences helps users structure and interpret their feelings (Brandtzæg et al., 2021). This function relates back to the expressive writing paradigm, where narrating emotional events supports cognitive and emotional processing (Pennebaker & Chung, 2007). Thus, chatbot dialogue can become a safe space to share private sentiments, particularly when humans are unavailable. Chatbots also offer interactive dialogue— such as asking follow-up questions—that traditional writing simply cannot.

Fourth, social ties also provide intellectual stimulation and entertainment: new ideas, alternative viewpoints, shared humor and play (Wright, 1997). Many users describe chatbot interactions as thought-provoking, playful, or creatively engaging, offering novel perspectives and exploration. This function is often appreciated particularly because chatbots do not impose social judgment (Rupprechter & Dienlin, 2025; Zhang et al., 2025).

Taken together, these functions clarify why relationships with artificial agents can become meaningful and what users stand to gain from them.

**Building and Maintaining Bonds**

Over time, emerging relational patterns and the functions they serve can solidify into stable attachment when an interaction partner provides emotional security and a reliable safe haven (Bowlby, 1974). Rewarding exchanges strengthen attachment and motivate its maintenance (Johnson & Rusbult, 1989). Similar bonding dynamics appear in human–chatbot interaction: users who experience emotionally validating, socially responsive exchanges report increased attachment and greater satisfaction (Xie & Pentina, 2022). By contrast, interactions remaining purely functional, potentially confined to machine talk, may rarely evoke emotional investment.

Attachment-relevant communication emerges when users treat the chatbot as a sophisticated social actor and relate to it through social talk. Repeated interaction and self-disclosure then establish relational routines that mirror attachment processes in human relationships (Pentina et al., 2023; Rupprechter & Dienlin, 2025; Skjuve et al., 2023). Although higher interaction frequency predicts stronger social connectedness to a chatbot (Christoforakos et al., 2021), frequency does not solely determine closeness: some users might interact often without engaging deeply, whereas brief but emotionally resonant exchanges can still feel meaningful (Skjuve et al., 2023), as discussed in the previous Communication chapter.

To understand how user-chatbot bonds are maintained over time, social exchange theory offers a useful lens (Cook et al., 2013). Each interaction prompts users to weigh rewards such as emotional support or stimulation against perceived costs such as technological breakdowns or privacy concerns (Krämer et al., 2011; Skjuve et al., 2023). Over time, repeated rewarding communication reinforces attachment (Pentina et al., 2023), whereas frustration or unmet expectations may weaken it (Konijn et al., 2025; Skjuve et al., 2023).

**Typology of Relationships**

After establishing how chatbot relationships form and what functions they provide, we can consider the types of relationships users develop with modern chatbots. Relational dimensions—role, mutuality, closeness, and interaction frequency—traditionally distinguish human bonds (Shaver & Reis, 1988). They span strong (high intimacy, frequent contact, sustained commitment) and weak ties (low intimacy, infrequent contact) (Granovetter, 1973). They can also take special forms. These include fanships (Stever, 2009) and negatively valenced relationships (e.g., enemyships). Chatbot relationships map onto these classifications in similar ways: some users relate to their chatbot as an acquaintance (Zhang et al., 2025), others form deep, attachment-like bonds resembling close friendships or romantic ties (Pentina et al., 2023). Users can also engage with character chatbots, typically designed to portray celebrities or fictional characters (e.g., Character.AI). In doing so, they extend fanships into otherwise unlikely contingent exchanges. Some users even engage in antagonistic relationships, bullying or trolling their chatbot (Rupprechter & Dienlin, 2025).

Crucially, all these relationships unfold in an inherently dualistic manner. Although chatbots lack lived experience, genuine self-disclosure (as there is, arguably, no underlying "self"), emotional intent, and currently the capacity to do otherwise, users nonetheless infer reciprocal intimacy and care, creating an *illusion* of mutuality (Horton & Wohl, 1956). Acknowledging this asymmetry of mutuality does not pathologize users' experiences. Parasocial scholars have long emphasized that such attachments can be equally meaningful (Giles, 2002). Consistent with our account of social perception, users often adopt a dual consciousness: they recognize that the chatbot cannot truly care, yet still maintain an emotional connection (Turkle, 2024). By partially overlooking the chatbot's artificiality, users can then fulfill emotional needs and "maximize the pleasure and emotional support one can derive from the 'friendship'" (E.-J. Lee, 2024, p.185).

In sum, attachment and companionship represent the potential outcomes of ongoing communication that holds both psychological promise and ethical complexity.

## Discussion

In this article, we introduced the AI relationship process (AI–RP) framework to capture how relationships with modern chatbots emerge. Drawing on the S–O–R–C model (Kanfer & Goldstein, 1991) and theories from communication, psychology, and HCI, the AI–RP specifies a sequential process linking four stages: (1) chatbot characteristics (modelled via a six-features profile), (2) users' social perceptions via dual routes, (3) communicative behavior modeled as a formative construct (breadth, depth, frequency, quality), and (4) relational outcomes such as attachment and companionship.

The AI–RP advances the field in three key ways. First, it addresses a central gap in current research: the absence of a *sequential*, *behavior*-focused framework of how human–chatbot relationships emerge. Our framework provides a conceptually rich, process-oriented integration of stimulus characteristics, user perceptions, communication behavior, and relational outcomes into a unified theoretical framework. As part of this contribution, the AI–RP is explicitly positioned in dialogue with existing theory.

Specifically, the framework aligns with emerging machine-bonding perspectives, such as affective bonding theory (Konijn et al., 2025), which explains why users form emotional ties with artificial partners once interaction becomes rewarding. It identifies antecedents of bonding, including perceived human-likeness, need-fulfillment potential, emotion-driven realism, and task-aligned affordances, and argues that bonds emerge when interaction yields valued rewards (Konijn et al., 2025). Together, the AI–RP and affective bonding theory clarify complementary aspects of artificial companionship: the AI–RP specifies the *observable communicative process* through which attachment can emerge, whereas affective bonding theory explains *motivational and perceptual aspects* that sustain it.

Beyond machine-bonding theories, the AI–RP opens avenues for integration with established communication models. Given the volume and often intimate nature of information exchanged with chatbots, particularly along the depth dimension, privacy research is likely to become an integral part of chatbot communication research. In particular, the AI–RP aligns closely with privacy models such as the privacy process model (Dienlin, 2014), which links privacy context, privacy perceptions, and communication behavior.

Second, the AI–RP foregrounds *communication* as the central user behavior and treats it as the primary stage through which human–chatbot relationships form. This approach remains underdeveloped in prior research (Konijn et al., 2025; Rapp, 2021). The AI–RP builds directly on its conceptual claims by introducing a *behavioral operationalization* of human-chatbot communication, shifting focus toward users' observable interaction behavior. In doing so, it addresses persistent conceptual ambiguities in human–chatbot research, including the "jingle problem" (Dibble et al., 2016), the "interactivity problem" (Hartmann, 2023), and the field's reliance on self-reports of interaction experience (e.g., Pentina et al., 2023).

By profiling chatbots as moderately realistic agents encountered in highly interactive settings comparable to human computer-mediated communication, the AI–RP reveals a crucial conceptual mismatch in existing research. Despite interactive conditions, human–chatbot communication is often examined through concepts of parasocial relationship theory (Horton & Wohl, 1956), which treats parasocial interaction as users' tendency to experience exchanges with artificial personas as reciprocal (Konijn et al., 2008) and has been extended to explain outcomes such as attachment, continued use, and engagement (Peng et al., 2025; Pentina et al., 2023; Tsai et al., 2021; Youn & Jin, 2021). However, the AI–RP clarifies that chatbot communication does not occur under the structural conditions parasocial interaction traditionally assumes—most notably one-sidedness (Hartmann, 2023).

One consequence of this mismatch is the widespread reliance on conceptually inconsistent measures that conflate parasocial interaction (short-term, one-sided exchange) with parasocial relationships (enduring bonds) (Dibble et al., 2016). Parasocial theory scholars have long emphasized that these constructs are distinct (Dibble et al., 2016). This represents a jingle fallacy (Hanfstingl et al., 2025), in which different concepts are mixed up and treated as identical.

A related issue concerns the "interactivity problem" (Hartmann, 2023). Because chatbots are encountered in genuinely interactive chat settings, rather than asymmetrical parasocial exchange (Dibble et al., 2016), chatbot communication is more appropriately understood as human–computer interaction. Consequently, CMC theories, such as social information processing (Walther, 2015) and the hyperpersonal model (Walther, 1996), arguably offer a more suitable analytical foundation. Yet, despite early work examining chat behavior directly (Corti & Gillespie, 2016; Park et al., 2018), systematic behavioral investigation remains limited (Rapp et al., 2021).

Finally, these conceptual issues are compounded by the field's reliance on self-report scales or qualitative accounts of interaction experience (Brandtzæg et al., 2022; Croes & Antheunis, 2021; Pentina et al., 2023; Skjuve et al., 2021). While such approaches provide indispensable insight into how users *feel* about chatbot interaction, they reveal little about what users actually *do* during communication. As a result, the field lacks measures that capture the genuinely reciprocal and behavioral nature of human–chatbot communication.

The implications of these issues become evident at the level of measurement. For example, the "AI Social Interaction Scale" includes items such as "Interactions with my Replika are reciprocal and mutual" alongside items such as "My Replika is like a friend." The former captures reciprocal exchange, whereas the latter reflects an existing relational bond. When such measures are used to predict downstream outcomes like attachment, effect sizes can become implausibly large (e.g., $\beta = .74$; Pentina et al., 2023, p. 9), because interaction and relationship are effectively measured twice (for similar cases; see Peng et al. 2025; Youn and Jin, 2021).

The AI–RP addresses these challenges directly. By clearly distinguishing communicative behavior from relational outcomes, acknowledging the genuinely bidirectional nature of human–chatbot interaction, and advancing a behavioral alternative to self-reports, the framework offers a precise conceptual and methodological foundation for studying chatbot communication. Supporting this approach, prior work shows that behaviorally grounded indicators substantially reduce inflated associations between interaction and attachment, yielding medium-sized effects (.20) rather than implausibly large ones (.73) (Rupprechter, 2024; Rupprechter & Dienlin, 2025). Such effect sizes appear more in line with established media effect sizes (Valkenburg et al., 2016).

Third, the AI–RP underscores the ethical significance of accurate measurement. Overestimated effects risk exaggerating claims about human–chatbot relationships and may contribute to moral panic surrounding them (Orben, 2020). Because the consequences of chatbot companionship remain ambivalent and emerging, accurate conceptualization of *how* they form is not merely a methodological concern but an ethical imperative. Empirical evidence already highlights both the social and emotional relevance of chatbot relationships. Users describe chatbots as confidants, friends, or romantic partners (Brandtzæg et al., 2022; Croes & Antheunis, 2021; Pentina et al., 2023), and some attribute major life changes to them (Rupprechter & Dienlin, 2025). While chatbot companionship can alleviate loneliness, provide emotional support, and foster transferable social skills (Brandtzæg & Følstad, 2017; De Freitas et al., 2025; M. Kim et al., 2025; Rupprechter & Dienlin, 2025), it also risks emotional dependency, unrealistic expectations, social isolation, and uncertain effects on well-being (Fang et al., 2025; Zhang et al., 2025). These are stakes particularly crucial for vulnerable populations, such as adolescent, neurodivergent, or lonely users.

**Limitations and Future Directions**

The AI–RP integrates multiple theoretical perspectives, but the framework is not exhaustive. Future work should test its components empirically, refine its sequential assumptions, and examine potential moderators (e.g., personality, social needs, or situative context). For instance, the proposed six-features profile captures core dimensions of chatbot characteristics while remaining extendable. Moreover, the behavioral construct of human–chatbot communication requires further validation through multimodal data (e.g., chat logs, longitudinal interaction patterns). Next, the AI–RP is a framework, outlining the core variables and mechanisms. It is not a *model*, aiming to incorporate *all* contributing variables and mechanisms. Specific paths amongst subdimensions, or variables not discussed here, might plausibly exist. Similar to the reinforcing spirals logic, also reciprocal effects could be modelled. For example, communication will also affect social perception, while growing attachment might also affect social perception and communication. Finally, future research should explore how evolving human-like design choices of modern companion chatbots, such as increasingly sycophantic systems, shape both communication and attachment formation.

**Conclusion**

In trying to make sense of the remarkable phenomenon of human–chatbot relationships, we find that the processes involved are both intricate and deeply human. Relationship building, for humans, is at once ancient and entirely new: a practice re-imagined through interaction with machines. Understanding the communicative behaviors that shape how we relate to AI chatbots not only informs our view of these emerging relationships but also reveals something fundamental about how we communicate with each other. We hope that the AI–RP offers a useful foundation for studying human–chatbot companionship as both a communicative and a social phenomenon—one that makes what users *do* an integral point of understanding.